\documentclass[%
 reprint,
 showpacs,
 amsmath,amssymb,
 aps,
 prl,
]{revtex4-1}

\usepackage{graphicx}
\usepackage{dcolumn}
\usepackage{bm}

\newcommand{\EQ}{\begin{equation}}
\newcommand{\EN}{\end{equation}}
\newcommand{\EQA}{\begin{eqnarray}}
\newcommand{\ENA}{\end{eqnarray}}

\graphicspath{{figures/}}

\begin{document}

\title{Temporal decorrelations in compressible isotropic turbulence}
\author{Dong Li}
\author{Xing Zhang}
\author{Guowei He}
\email{hgw@lnm.imech.ac.cn}
\affiliation{LNM, Institute of Mechanics, Chinese Academy of Sciences, Beijing, 100190, China}


\begin{abstract}

Temporal decorrelations in compressible isotropic turbulence are studied using the space-time
correlation theory and direct numerical simulation.  A swept-wave model is developed for
dilatational components while the classic random sweeping model is proposed for solenoidal components.
The swept-wave model shows that the temporal decorrelations in dilatational fluctuations are dominated
by two physical processes: random sweeping and wave propagation.
These models are supported by the direct numerical simulation of compressible isotropic turbulence,
in the sense of that all curves of normalized time correlations for different wavenumbers collapse
into a single one using the normalized time separations.
The swept-wave model is further extended to account for a constant mean velocity.
\end{abstract}



\maketitle



A milestone in isotropic and homogeneous turbulence is the
random sweeping hypothesis \cite{kraichnan1964,tennekes1975}.
The random sweeping hypothesis proposes that there is a temporal decorrelation
process in incompressible isotropic turbulence and develops a simple
model for space-time correlations of velocity fluctuations \cite{kraichnan1964}.
These results are examined theoretically \cite{yakhot1989,chensy1989,nelkin1990}
and verified experimentally \cite{praskovsky1993} and numerically \cite{sanada1992,he2004}.
The space-time correlation models are used to predict the scalings
of wavenumber or frequency energy spectra in turbulent flows \cite{rubinstein1999, sagaut2008}.
The decorrelation processes are also relevant to the non-Gaussian statistics \cite{kaneda1999}
and intermittency \cite{tsinober2001}. Their further applications can be found in
turbulence generated noise \cite{wang2006}.
The recently increasing studies on compressible isotropic turbulence raise such a question
on the effects of compressibility on decorrelation processes \cite{benzi2008,pan2009,aluie2011}.
In this letter, we will study the decorrelation processes in compressible isotropic turbulence
and propose a model for space-time correlations of dilatational components.

A compressible turbulence is associated with two characteristic velocities: fluid velocity
and sound speed, whereas an incompressible one is only associated with
fluid velocity. Therefore, the decorrelation processes in compressible turbulence are very
different from incompressible one. A space-time correlation is the essential
quantity to measure the decorrelation processes in turbulent flows. Three typical models exist
for space-time correlations in turbulence theory. The first one is, as stated
above, the random sweeping model for incompressible turbulence \cite{kraichnan1964}.
We will show that it cannot characterize the acoustic components in compressible turbulence.
The second one is the Taylor frozen flow model \cite{taylor1938}. It has been shown
that this model is not a good approximation for dilatational components \cite{lee1992}.
The third one is the linear wave propagation model \cite{lee1992}. This model can be
used for dilatational components if compressible turbulence has a dominating mean velocity.
However, we will find that it does not decrease with increasing temporal separation, which violates
the nature of correlation functions.

In the present letter, we will develop a space-time correlation model for compressible isotropic
turbulence. This is achieved by the Helmholtz decomposition: a velocity field can be split into
the solenoidal and dilatational components. A swept-wave model will be developed for the dilatational
components while the solenoidal components are expected to follow the random sweeping model.
The swept-wave model will be numerically validated and further used to elucidate the decorrelation
process in compressible turbulence.

We consider compressible and isotropic turbulence with periodic boundary conditions.
In this case, the Helmholtz decomposition for velocity fields can be made as follows \cite{sagaut2008}
\EQ
{\bf u} = {\bf u}^s + {\bf u}^d,
\label{decomposition}
\EN
where ${\bf u}^s$ and  ${\bf u}^d$ are the solenoidal (i.e. incompressible) and
dilatational components, respectively. The harmonic component is taken to be zero.
We will investigate the temporal decorrelations of solenoidal and dilatational
components. The temporal decorrelations can be measured by the space-time correlations
of velocity fluctuations
\EQ
R(r,\tau) =\langle u_{\ell}( {\bf x}, t) u_{\ell} ( {\bf x} + {\bf r}, t+\tau) \rangle,
\EN
or its equivalent forms in Fourier space
\EQ
{\hat R}(k,\tau) =\langle \hat{u}_{\ell}( {\bf k}, t) \hat{u}_{\ell} ( -{\bf k}, t+\tau) \rangle.
\EN
Here, r and k are the magnitudes of separation vector ${\bf r}$ and wavenumber vector ${\bf k}$.
The similar quantities $R^s$ and $R^d$ can be defined for the solenoidal and dilatational
components ${\bf u}^s$ and ${\bf u}^d$, respectively.

We propose that a solenoidal component follows the same decorrelation process as
the random sweeping process for incompressible isotropic turbulence: small eddies
are randomly convected or swept by energy-contained eddies, where the contribution
of dilatational components to the energy-contained eddies is comparably small \cite{moin2009}.
The random sweeping process can be described by a simple idealized convection
equation as follows \cite{kraichnan1964}
\begin{equation}
\left( \frac{\partial}{\partial t} + v_j \frac{\partial}{\partial x_j} \right) u^s_{\ell} = 0,
\label{sweep-equ}
\end{equation}
\noindent
where ${\bf v}=(v_1, v_2, v_3)$  is a spatially uniform and stationary Gaussian
random field. $V=|{\bf v}|/\sqrt{3}$ is the r.m.s of solenoidal components along one axis.
Its solution in the Fourier space is given by

\begin{equation}
\hat{u}^s_{\ell}({\bf k}, t)
= \hat{u}^s_{\ell} ({\bf k}, 0) \exp\left[- i ({\bf k \cdot v}) t \right].
\label{sweep-solu}
\end{equation}

\noindent
The time correlation of Fourier mode is formulated as

\begin{equation}
{\hat R}^s(k, \tau)= {\hat R}^s(k,0) \exp\left(-\frac{1} {2} V^2 k^2 \tau^2 \right).
\label{sweep-corr}
\end{equation}

\begin{figure}
\includegraphics[width = 0.45\textwidth]{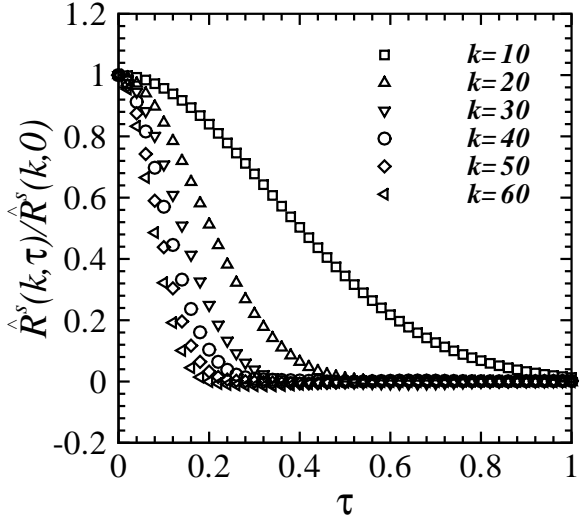}
\caption{\label{fig:us} Normalized time correlations of solenoidal components vs time separations for $k=10,20,30,40,50,60$.}
\end{figure}

\begin{figure}
\includegraphics[width = 0.45\textwidth]{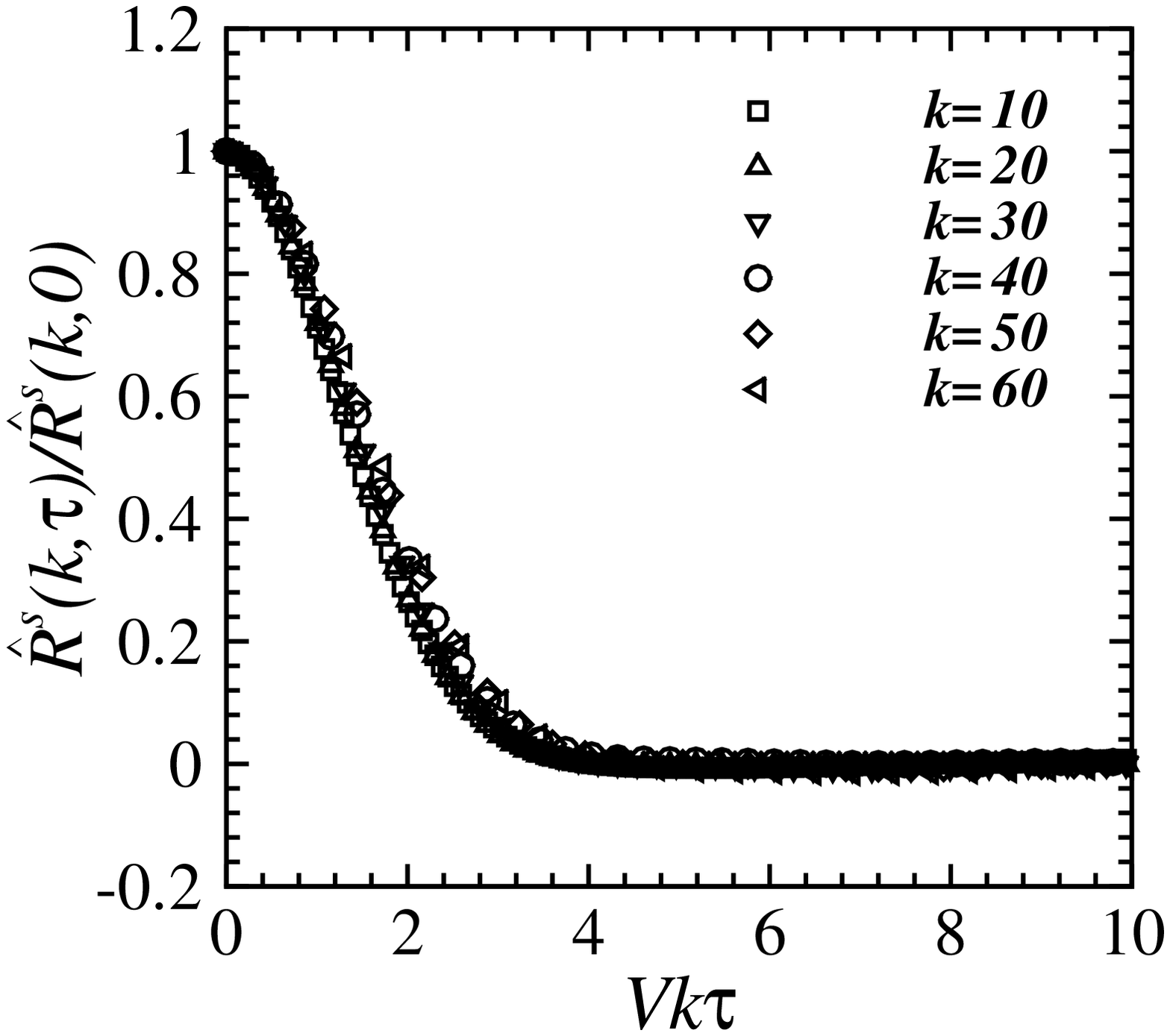}
\caption{\label{fig:usk} Normalized time correlations of solenoidal components vs normalized time separations
$V k \tau$ for $k=10,20,30,40,50,60$.}
\end{figure}

A dilatational component in compressible isotropic turbulence propagates at the speed of
sound relative to moving fluids.
This implies that the dilatational fluctuations are swept by the energy-contained eddies.
Therefore, the temporal decorrelations in dilatational components are governed by two dynamic processes:
random sweeping and wave propagation. The well-known linear wave propagation model \cite{lee1992}
only includes the wave propagation process. In order to account for the random sweeping
effect, we introduce a new term ${\bf v \nabla}$ into the linear wave propagation
equation and propose the governing equation for dilatational fluctuations as follows
\begin{equation}
\left( \frac{D^2}{D t^2} - \bar{a}^2 \nabla^2 \right) u^d_{\ell} = 0,
\label{wave-equ}
\end{equation}
where ${\bar a}$ is the mean speed of sound and
\begin{equation}
\frac{D}{Dt} = \left( \frac{\partial}{\partial t} + {\bf v} \cdot \nabla \right)
=  \left( \frac{\partial}{\partial t} + v_j \frac{\partial}{\partial x_j} \right).
\end{equation}
Here, ${\bf v}$ is the same Gaussian random field as one in Eq.~(\ref{sweep-equ}).
The new term ${\bf v \nabla}$ in Eq.~(\ref{wave-equ}) represents the random
sweeping effect, which is absent in the linear wave propagation model \cite{lee1992}.
The solution of Eq.~(\ref{wave-equ}) in Fourier space is given by
\begin{eqnarray}
\hat{u}^{d}_{\ell} ({\bf k},t)&=& \hat{u}^{d+}_{\ell}({\bf k},0) \exp \left[ -i ({\bf k \cdot v}) t
- i k \bar{a} t \right] +
\nonumber \\
& & \hat{u}^{d-}_{\ell}( {\bf k},0) \exp \left[ -i ({\bf k \cdot v}) t + i k \bar{a} t \right],
\label{wave-solu}
\end{eqnarray}
\noindent
where $\hat{u}^{d+}_{\ell}$ and $\hat{u}^{d-}_{\ell}$ are the Fourier coefficients.

The time correlation of Fourier mode is calculated as follows
\begin{eqnarray}
{\hat R}^d(k,\tau) &=& \langle \hat{u}^d_{\ell}({\bf k}, t) \hat{u}^d_{\ell}(-{\bf k}, t+\tau)\rangle \label{rc} \\
&=& \langle \hat{u}^{d+}_{\ell}({\bf k},0) \hat{u}^{d+}_{\ell}(-{\bf{k}},0) \rangle
\langle \exp[{i ({\bf k \cdot v})\tau} + i k \bar{a}\tau] \rangle \nonumber \\
&+& \langle \hat{u}^{d-}_{\ell}({\bf k},0) \hat{u}^{d-}_{\ell}(-{\bf{k}},0) \rangle
\langle \exp[{i ({\bf k \cdot v})\tau} - i k \bar{a}\tau] \rangle. \nonumber
\end{eqnarray}
Here, the mode correlation $ R^d(k,0) $ is given by
\begin{equation}
\frac{1}{2} {\hat R}^d(k,0) =
\langle{\hat{u}^{d+}_{\ell}(-\mathbf{k},0) \hat{u}^{d+}_{\ell}(\mathbf{k},0)}\rangle
= \langle{\hat{u}^{d-}_{\ell}(-\mathbf{k},0) \hat{u}^{d-}_{\ell}(\mathbf{k},0)}\rangle.
\nonumber
\end{equation}
Therefore, the correlation functions can be expressed as
\begin{equation}
\displaystyle\frac{ {\hat R}^d(k,\tau)}{{\hat R}^d(k,0)}
=\cos(k\bar{a}\tau) \exp\left[-\frac{1}{2} k^2 V^2 \tau^2 \right].
\label{swept-wave}
\end{equation}
The swept-wave model (\ref{swept-wave}) contains two factors: a linear wave function
and an exponential function.
The first factor represents the random sweeping effect and the second one represents
the wave propagation process. If the sweeping velocity is zero, it becomes the linear
wave propagation model. In fact, the linear wave propagation model in compressible
isotropic turbulence is simplified as $\cos(k\bar{a}\tau)$, which is the inverse Fourier
transformation of Equation (19) in \cite{lee1992}. This cosine function does not decay
to zero as time separation increases.

To validate the swept-wave model, we solve the three-dimensional, compressible Navier-Stokes equations in a cubic box of side $2\pi$ with specific heat ratio $\gamma=1.4$ and Prandtl number $Pr=0.7$ using an optimized sixth order compact, finite difference scheme. Statistically stationary flow fields are achieved by including a forcing term $\mathbf{f}(\mathbf{x},t)$ with solenoidal modes only, $\nabla\cdot\mathbf{f}=0$. The forcing term is nonzero only in the range $1 \leq k \leq 3$ and obeys a Gaussian random distribution with an exponential temporal correlation. Each component of $\mathbf{\hat{f}}(\mathbf{k},t)$  is defined by $\hat{f}_i=(\delta_{ij}-k_i k_j / k^2)\hat{g}_j$, where $\hat{g}_j$ is generated by an independent Uhlenbeck-Ornstien process. After the flow fields become statistically stationary, a total of 400 flow fields with computational time increment $0.02$ are chosen to calculate time correlations. The Taylor's micro-scale based Reynolds number is about 80 and the turbulent Mach number is about 0.63. We also compare the space-time correlations from the present case with ones from decaying turbulence. The results obtained are consistent.

Fig.~\ref{fig:us} shows the correlation coefficients of solenoidal modes for wavenumbers
$k=10, 20, \cdots$ $50$ and $60$, spanning a range of scales from the integral scale
to the dissipation scales. The correlation coefficients are the normalized correlation functions
by the mode correlation ${\hat R}^s(k,0)$. Obviously, the solenoidal modes decorrelate more
quickly at larger wavenumbers than at small wavenumbers. These results in Fig.~\ref{fig:us} are all plotted together in Fig.~\ref{fig:usk}, with the horizontal axis defined by the normalized time scale $V k\tau$. This normalization causes excellent collapse of the correlation coefficients. The collapse on the normalized time scale $V k \tau$ supports the random sweeping hypothesis for the solenoidal components.

\begin{figure}
\includegraphics[width = 0.45\textwidth]{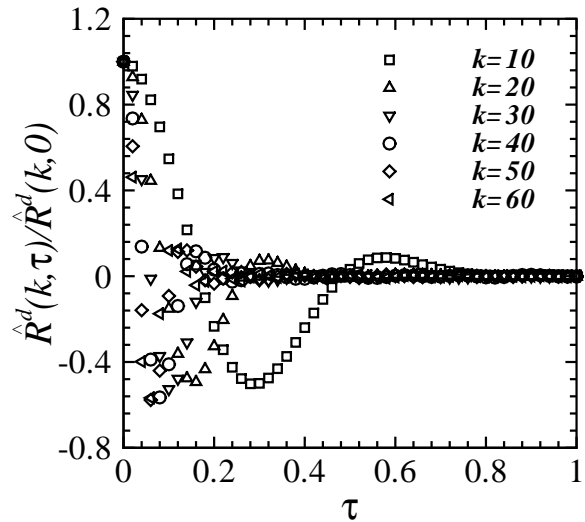}
\caption{\label{fig:uc} Normalized time correlations of dilatational components vs time separations for $k=10,20,30,40,50,60$.}
\end{figure}

\begin{figure}
\includegraphics[width = 0.45\textwidth]{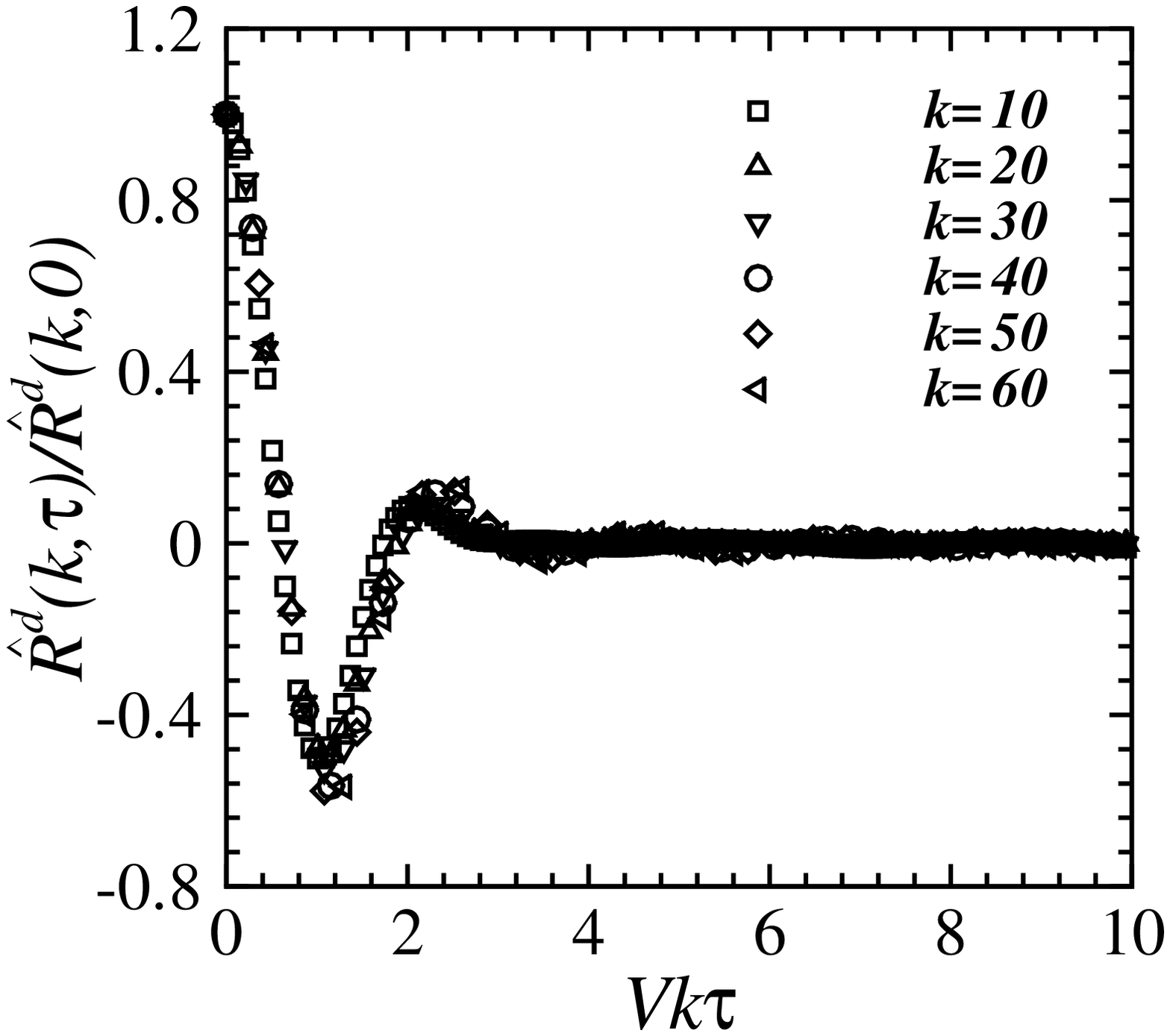}
\caption{\label{fig:uck} Normalized time correlations of dilatational components vs normalized time separations $V k \tau$ for $k=10,20,30,40,50,60$.}
\end{figure}

Fig.~\ref{fig:uc} plots the normalized time correlations ${\hat R}^d(k,\tau)/{\hat R}^d(k,0)$ for
dilatational components from DNS data for wavenumber $k=10, 20, \cdots$ and $60$, where the
correlations are normalized by the correlation ${\hat R}^d(k,0)$. It is observed from
Fig.~\ref{fig:uc} that the time correlations of dilatational components decay with oscillations.
This is very different from solenoidal components where the time correlations decay
without any oscillation. These oscillatory decays confirm that temporal decorrelations
in dilatational components are mainly determined by both random sweeping and wave propagation.

Fig.~\ref{fig:uck} presents the normalized time correlations ${\hat R}^d(k,\tau)/{\hat R}^d(k,0)$
versus the normalized time separation $Vk\tau$ for wavenumber $k=10, 20, \cdots$, $50$ and $60$.
The time normalization leads to the virtual collapse of all curves. The collapsed curves decay with oscillations. This result verifies the proposed swept-wave model for dilatational components. We note
that the ratio of two scale-similarity variables $Vk\tau$ and ${\bar a}k\tau$ is constant.

\begin{figure}
\includegraphics[width = 0.45\textwidth]{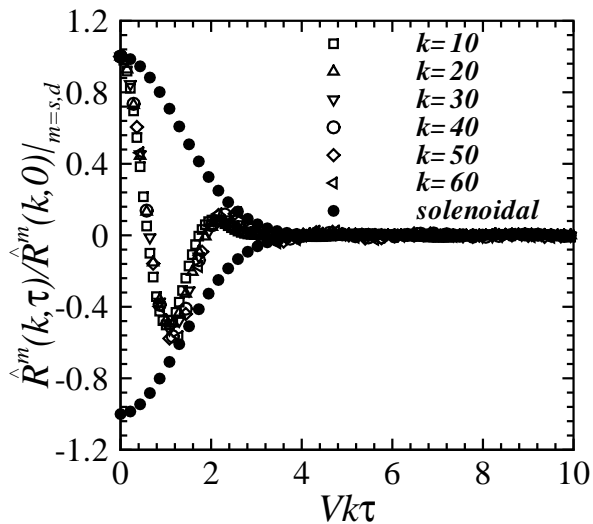}%
\caption{\label{fig:uc_us_usm_ucm} Normalized time correlations vs normalized time separations $V k \tau$.
Dilatational components for $k=10, 20, 30, 40, 50, 60$; solenoidal component and its mirror image for $k=30$.}
\end{figure}


Fig.~\ref{fig:uc_us_usm_ucm} compares the collapsed curves for solenoidal components with the collapsed
ones for dilatational components, where the horizontal axis is normalized as the scale-dependent similarity
variable $Vk\tau$. It is observed that the collapsed curves for solenoidal components act as an envelop
of the collapsed one for dilatational components. This confirms the swept-wave model where the exponential
function $\exp(-0.5 V^2 k^2 \tau^2)$ acts as an envelop.

We can further calculate the space-time correlations of dilatational components using
the Fourier transformation of ${\hat R}^d(k,\tau)$ from wavenumber space to spatial one
\begin{eqnarray}
R^d(r,\tau) &=& \int_0^\infty E(k) \exp \left(-\frac{1}{2}V^2 k^2 \tau^2 \right)
\nonumber \\
& & \cos(k \bar{a} \tau) \frac{\sin(k r)}{k r}  d k.
\label{sweptwave-phys}
\end{eqnarray}
Eq.~(\ref{sweptwave-phys}) can be extended to account for a constant mean velocity.
Without loss of generality,
we choose the constant mean velocity $U_1$ in the direction of the $x_1$ axis. Applying the
coordinate transformation $(y_1=x_1-U_1t, y_2=x_2, y_3=x_3)$ to Eq.~(\ref{wave-equ}), we obtain
\begin{eqnarray}
R^d(r,\tau) &=& \int_0^\infty E(k) \mathrm{exp}\left(-\frac{1}{2}V^2 k^2 \tau^2 \right)
\nonumber \\
& & \cos(k \bar{a} \tau) \frac{\sin \left[ k (r - U_1 \tau ) \right]}{k (r - U_1 \tau )}  d k.
\label{sweptwave-ConstVel}
\end{eqnarray}
In comparison with the linear wave propagation model, this model (\ref{sweptwave-ConstVel})
contains an additional exponential function that is responsible for the random sweeping effect.
It also confirms that Taylor's frozen flow model is not a good approximation to the
space-time correlation of dilatational components. Wilczek and Narita \cite{wilczek2012}
consider the random sweeping model with constant mean velocity. The present model
is consistent with their results.

In summary, we find that solenoidal and dilatational components in compressible isotropic
turbulence display different decorrelation processes: a dilatational component is dominated
by both random sweeping and wave propagation while a solenoidal component dominated
by the random sweeping effect. We further develop a swept-wave model for dilatational
fluctuations. This model is distinct from the linear wave propagation model since it
includes the random sweeping process.
The DNS data validates the swept-wave model for compressible isotropic turbulence.
The further extension of the swept-wave model to turbulent shear flows is referred
to the elliptic model \cite{he2009} and the present model can be used to study the propagation
velocity of coherent structures in compressible turbulence.

Acknowledgements: This work is supported by National Natural Science Foundation of China under projects
No.~11232011(Key project) and No.~11021262(Innovative team)and the National Basic Research Program of
China (973 Program) under Project No.~2013CB834100 (Nonlinear science).

\bibliography{[2013Jan10]compressible}

\end{document}